\documentclass[12pt, a4paper]{article}
\usepackage[utf8]{inputenc}
\usepackage{amsmath}
\usepackage{amssymb} 
\usepackage{amsthm}
\usepackage{physics}
\usepackage{authblk}
\newcommand{\sgn}{\operatorname{sgn}}
\title{Thermodynamical Properties of Bosons and Fermions for a Lattice Motivated Dispersion Relationship}
\author{Metin Güner\thanks{e-mail:metinmguner@yahoo.com}}
\author{Metin Arık\thanks{e-mail:metin.arik@boun.edu.tr}}
\affil{Physics Department, Boğaziçi University, 34342, Istanbul, Turkey}

\date{\today}
\begin{document}
	
\maketitle

\begin{abstract}
	In this paper we calculate the basic thermodynamical quantities for a system of bosonic simple harmonic oscillators (BSHOs) and the corresponding system of fermionic simple harmonic oscillators (FSHOs) using a dispersion relationship similar to cases in general relativity and condensed matter physics. In the FSHO we see negative temperatures in both cases we obtain finally a system of oscillators with only one effective frequency of vibration. Also we see that pressure is less than zero for $T<T_c$ where $T_c$ is a critical temperature for bosons. 
\end{abstract}

\section{Introduction}
\hspace{\parindent} In this paper we calculate the thermodynamic properties of a system of bosons and of a system of fermions obeying the dispersion relation $\omega^2=m^2+\Delta^2\sin^2(\pi k a/L)$ in units with $\hbar=1$, $c=1$. This dispersion relationship is similar to the cases in solid state physics as in \cite{ashcroft1976solid}, in the study of lattice vibrations in a solid. On the other hand for small a it becomes the dispersion relation for a free particle with momentum $k$ in 1+1 dimensional Minkowski space. The dispersion relationship above also arises \cite{arik2016scalar} if  space has a discrete structure with total length $L$ which is an integer multiple of  a with periodic or nonperiodic boundary conditions. 

In the literature there exists similar calculations although in a different physical content. For example, Unruh \cite{unruh1995sonic} has pointed out that one can employ certain dispersion relationships arises in the context of dumb holes: i.e. entities which can occur when the velocity of the fluid increases to values larger than the velocity of sound on a closed surface. The family of relationships he uses is given by 
\begin{equation}
\omega_\pm=v k \pm f^2(k)
\end{equation}
and
\begin{equation}
f^2(k)=k_0\left(\tanh\left(\frac{k}{k_o}\right)^n\right)^{1/n}.
\end{equation}
He reasoned that using this relationship he could apply it to the case of black holes. The physical aspects of this calculation is not so interesting for us. The important point is that such relationships do occur and thus can be utilized in the context of real physical problem. 

On the other hand Casadio \cite{casadio2002dispersion} has investigated the dependence of the black hole evaporation on the high frequency behavior of Hawking quanta in 4-dimensions, using a particular dispersion relationship. The relationship he used is (see also \cite{mersini2001relic})
\begin{equation}
\omega^2=\frac{\tilde{k}^2}{k_0^2}\left(\frac{\varepsilon_1}{(1+e^{\tilde{k}/k_0})}+\frac{4-2\varepsilon}{(1+e^{\tilde{k}/k_0})^2}\right).
\end{equation}

He obtained the result that even in the case of large deviations in the dispersion relationship in the ultra high frequency domain the luminosity of the black hole was not changed appreciably for a judicious choice of parameters.

In addition to the above Mersini et. al. \cite{mersini2001relic} focused on the impact the trans-Planckian regime has on the observables, i.e. dark energy and cosmic microwave background, radiation spectrum. The dispersion relationship they use 
\begin{equation}
\omega^2(k)=f^2(k)=k^2\left(\frac{\varepsilon}{1+e^x}+ \frac{\varepsilon_2e^x}{1+e^x}+\frac{\varepsilon_3e^x}{(1+e^x)^2}\right).
\end{equation}
They demand that dispersion relationship goes asymptotically to zero i.e. $\varepsilon_2=0$. Also the condition of a nearly linear dispersion relationship for $k<\tilde{k}$ requires that $\varepsilon_1/2+\varepsilon_3/4=1$.

Also Corley and Jacobson \cite{corley1996hawking} has used a relationship 
\begin{equation}
f^2(k)=
\begin{cases}
k^2-\frac{k^4}{k_0^2}&\qquad k^2<k_0^2\\
\quad 0&\qquad k^2>k_0^2.
\end{cases}
\end{equation}

This dispersion relationship has the same small $k$ behavior as Unruh's but behaves quite differently for large $k$. They focused on to qualitatively different types of particle creation, a thermal Hawking flux generated by a process called mode conversion at the black hole horizon, and a non-thermal spectrum caused by scattering from a stationary geometry.

Brout et. al. \cite{brout1995hawking}, has solved Unruh's problem analytically (for the dumb black holes) and they have obtained a similar result.

Dispersion relationships similar to the above form occur in condensed matter physics as well. For instance in the phenomena of plasma oscillations we have
\begin{equation}
\omega_{k}^2=\omega_{p}^2+c^2k^2.
\end{equation}
The similarity with our case is obvious for small $k$. Also for a system of harmonic oscillators on a lattice as in condensed matter physics we have that 
\begin{equation}
\omega^2=\frac{k_1+k_2}{2m}\left[1\pm\left[1-\frac{4k_1k_2}{(k_1+k_2)^2} \sin^2(ka/2)\right]^{1/2}\right]
\end{equation}
where $k_1$ and $k_2$ are the spring constants for a system with diatomic basis. Where + (-) sign corresponds to the optical (acoustical) mode in the case $4k_1k_2/(k_1+k_2)^2\ll 1$ irrespective of the value $k$. On the other hand for $k_1=k_2=k$ it is also similar to the case in this paper. Indeed  
\begin{equation}
\omega^2=\omega_\pm^2=\frac{k}{m}\left[1\pm \cos(\frac{ka}{2})\right]
\end{equation} 
for acoustical and optical phonon modes.

It is also possible to calculate the relevant quantities for the acoustical modes (work in progress). We will take $k$ to be a one dimensional continuous variable and use the density matrix and statistical mechanics to calculate the thermodynamic properties. This method is indeed well known for a system of harmonic oscillators which we briefly review in Section 2. In Section 3 we apply this dispersion relation to a system of bosonic oscillators and obtain infinite  compressibility as in Bose-Einstein condensation. We show that the pressure is negative for temperatures below a critical temperature. In Section 4 the same dispersion relation is used for a system of fermionic oscillators for which it is shown that negative temperatures exist. As temperature approaches zero from above or below, the system exhibits a discontinuous behaviour for energy and pressure.

One important point that we obtain for both the bosonic and the fermionic case is that the mass dependence of the thermodynamic variables  such as energy and entropy m is through an effective mass parameter $\omega_0=m(1+ \Delta^2/m^2)^{1/4}$ which shows that the zero mass limit may be singular. Section 5 is reserved for conclusions.

\section{The General Formalism for a System of Coupled Quantum Harmonic Oscillators}

\hspace{\parindent}The problem of the harmonic oscillator is probably the simplest and most widely used amongst all the non-trivial cases in quantum mechanics. The kernel (propagator) for a single oscillator is given by 
\begin{equation}
K(x,t;x',t')=\left(\frac{m\omega}{2\pi i h \sin(\omega T)}\right)^{1/2}\exp(\frac{i m \omega}{2 h\sin(\omega T)}((x^2+x'^2)\cos(\omega T)-2xx'))
\end{equation}
where $T=t-t'$ and the other variables are as widely used in literature. A general form for the kernel is as follows
\begin{equation}
K(x,t;x',t')=\sum_{j}  \phi_j^*(x) \phi_j(x') e^{-i(t_2-t_1)E_j}.
\end{equation}

Now we go to imaginary time ($t=i\beta$) i.e. to the realm of statistical mechanics. A closely related expression is  
\begin{equation}
\rho(x,x')=\sum_{j} \phi_j^*(x) \phi_j(x') e^{-\beta E_j}
\end{equation} 
where $\rho(x,x')$ is the density matrix. Now take the trace of the density matrix i.e. take the integral over $x$ after setting $x=x'$
\begin{equation}
\begin{split}
\int dx \rho(x,x')&=\sum_{j} \left(\int dx \phi_j^*(x) \phi_j(x)\right)  e^{-\beta E_j}\\
&=\sum_{j}e^{-\beta E_j}=Z\quad\text{since}\quad \int dx \phi_i^*(x) \phi_j(x)=\delta_{ij}.
\end{split}
\end{equation}

Thus to calculate the partition function we can use $\rho(x,x')$. Indeed the differential equation for our kernel (in the quantum mechanical case) is given by 
\begin{equation}
-\frac{h}{i}\frac{\partial K(2,1)}{\partial t}=H_2K(2,1)
\end{equation} 
where $H_2$ acts on the variable denoted by $2$. 

For our statistical mechanical system on the other hand 
\begin{equation}
-\frac{\partial \rho(2,1)}{\partial \beta}=H_2 \rho(2,1)
\end{equation}
where $H_2$ acts on $2$ again.

Thus for example we can calculate $K$, the kernel in our quantum mechanical case using path integrals and then analytically continue to imaginary time where $\beta=i (t_2-t_1)$ to obtain $\rho(x,x')$ and from there calculate the partition function by taking the trace over $x$.

Of course now we can extend our physical system of a single harmonic oscillator into a system of coupled (for the moment) harmonic oscillators. We again use imaginary time and employ the Euclidean Lagrangian. For our system of coupled harmonic oscillators we have 
\begin{equation}
L=\frac{1}{2}\sum_{j} \dot{q}_j^2+\frac{1}{2} \sum_{j,k=1} V_{jk} q_jq_k.
\end{equation}

We make an orthogonal transformation to a state which consists of independent oscillators. Our Lagrangian now reads
\begin{equation}
L=\frac{1}{2}\sum_{j}\left( \dot{Q}_j^2+\omega_j^2 Q_j^2 \right)
\end{equation}
where $Q_j$ are the new position variables, $\omega_j$ is the eigenfrequency corresponding to the mode denoted by $j$.

Now we have a set of uncoupled harmonic oscillators. For a single oscillator we have for the partition function $Z=-\log(2\sinh(\frac{\beta \omega}{2}))$ for a bosonic simple harmonic oscillator (BSHO) and $Z=\log(2\cosh(\frac{\beta \omega}{2}))$ for a fermionic simple harmonic oscillator (FSHO).

For a collection of independent oscillators we have 
\begin{equation}
Z=\prod_k Z_k.
\end{equation}

After calculating this function which is closely related to Helmholtz free energy $A$ we can easily obtain the thermodynamic quantities $A$, $E$, $S$, $P$, $C_L$ respectively Helmholtz free energy, energy, entropy, pressure, specific heat at constant volume. For the particular dispersion relationship we choose:
\begin{equation}
\omega_k^2=m^2+\Delta^2 \sin[2](\frac{k \pi a}{L})
\end{equation}
where $k$ is an integer, $k=0,1,\dots,L/a-1$ and $L$ stands for the so called volume, and $a$ stands for the spacing of the lattice (or merely inserted for dimensional consistency). We will go to the continuous limit for solving this equation.

As in the introduction dispersion relationships of this kind are encountered in different context in general relativity, condensed matter physics and also in discrete space structures \cite{arik2016scalar} when the total length $L$ is an integer multiple of $a$, the lattice spacing. In addition to our calculation is valid for all $\Delta/m\ll 1$ but for all $T$.

We note finally that we solved this model for both bosonic and fermionic cases and separately obtained similar results. 

\section{Calculation of the Partition Function of Bosonic Oscillators}
\subsection{Preliminaries} 
\hspace{\parindent}First we make the passage from the discrete case to the continuous case and determine exactly what we are going to calculate. From the above equations again since 
\begin{equation}
\omega^2_k=m^2+ \Delta^2 \sin[2](\frac{\pi k a}{L}),\qquad 0\leq k\leq \frac{L}{a}
\end{equation}
and
\begin{equation}
\log(Z)=\sum_{k}\log(Z_k)=-\sum_{k} \log(2\sinh(\frac{\beta \omega_k}{2}))
\end{equation}
by enumeration of momentum states
\begin{equation}
\Delta k=\frac{L}{\pi a} \frac{\omega \Delta \omega}{\sqrt{(\omega^2-m^2)(\Delta^2+m^2-\omega^2)}}
\end{equation}
we have
\begin{equation}
\log(Z)=\frac{-L}{\pi a}\int_{m}^{\sqrt{m^2+\Delta^2}}\frac{\omega d\omega}{\sqrt{(\omega^2-m^2)(\Delta^2+m^2-\omega^2)}}\log(2\sinh(\frac{\beta \omega}{2})).
\end{equation}
Noting that
\begin{align}
\log(2\sinh(\frac{\beta \omega}{2}))&=\frac{\beta \omega}{2}+\log(1-e^{-\beta \omega}) \notag \\
&=\frac{\beta \omega}{2}-\sum_{n} \frac{e^{-n\beta \omega}}{n}
\end{align}
the second term on the right is the Bose-Einstein integral function of order one. We refer the reader to the literature \cite{pathria2011statistical} and merely give the result here. In the notation of \cite{pathria2011statistical},
\begin{align}
g_1(e^{-\beta \omega}) &= \sum_{n} \frac{e^{- \beta \omega}}{n} \notag \\
&= -\log(\beta \omega)+ \sum_{n=1}^{\infty} \frac{(-1)^n \zeta (1-n) (\beta \omega)^n}{n!}.
\end{align}
Here $\zeta(z)$ is the Riemann zeta function. Since $\zeta(-2n)=0$ only $\zeta(1-2n)$ and $\zeta(0)$ terms contribute and we have the expression \cite{whittaker1996course} 
according to $\zeta(1-2n)=\frac{-1^nB_n}{2n}$ where $B_n$ are Bernoulli numbers and $\zeta(0)=-1/2$. As a result 
\begin{equation}
g_1(e^{-\beta \omega})=-\log(\beta \omega)+\sum_{n=1}^{\infty} \frac{(-1)^n B_n (\beta \omega)^{2n}}{(2n)!2n}+\frac{\beta\omega}{2}.
\end{equation}
Therefore we have,
\begin{equation}\label{eq8}
\log(2\sinh(\frac{\beta \omega}{2}))= -\sum_{n=1}^{\infty} \frac{(-1)^n B_n (\beta \omega)^{2n}}{(2n)!2n}+\log(\beta \omega).
\end{equation}

\subsection{Calculation of the Partition Function }
\hspace{\parindent}We will now seperately calculate the  contributions to the partition function from the first and second terms of (\ref{eq8}). Hence we set 
\begin{equation}
-\log(Z)=I+K
\end{equation}
where
\begin{equation}\label{eq20}
I=-\frac{L}{\pi a}\sum_{n=1}^{\infty} \frac{(-1)^n B_n \beta^{2n}}{(2n)! 2n} \int_{m}^{\sqrt{m^2+\Delta^2}}\frac{\omega^{2n+1} d\omega}{\sqrt{(\omega^2-m^2)(\Delta^2+m^2-\omega^2)}}
\end{equation}
and
\begin{equation}\label{eq21}
K=\frac{L}{\pi a} \int_{m}^{\sqrt{m^2+\Delta^2}} \frac{\omega \log(\beta \omega)d\omega}{((\omega^2-m^2)(\Delta^2+m^2-\omega^2))^{1/2}}.
\end{equation}
Now defining for the first term of (\ref{eq8}),
\begin{equation}I_n=\int_{m}^{\sqrt{m^2+\Delta^2}}\frac{\omega^{2n+1} d\omega}{\sqrt{(\omega^2-m^2)(\Delta^2+m^2-\omega^2)}}
\end{equation}
we get
\begin{equation}
I=-\frac{L}{\pi a} \sum_{n=1}^{\infty} \frac{(-1)^n B_n}{(2n)!2n} \beta^{2n} I_n.
\end{equation}
Setting $\omega^2=m^2+z$,
\begin{equation}\label{eq22}
I_n=\frac{1}{2} \int_{0}^{\Delta^2} \frac{(m^2+z)^n dz}{(z (\Delta^2-z))^{1/2} }.
\end{equation}
Now from Appendix A we have,
\begin{equation} \label{eq23}
I_n= \frac{1}{2} m^{2n}\, _{2}F_{1} (-n, 1/2;1,-\Delta^2/m^2)B(1/2,1/2)
\end{equation} 
where $_{2}F_{1}$ is the hypergeometric function of its arguments and $B$ is the beta function.\par
Now we use Appendix B and express the hypergeometric function in terms of Legendre polynomials. Thus we have
\begin{equation}\label{eq26}
_{2}F_{1} (-n, 1/2;1,-\Delta^2/m^2)=(1+\frac{\Delta^2}{m^2})^{n/2} P_{-n-1}\left( \frac{1+\frac{\Delta^2}{2m^2}}{\left(1+ \frac{\Delta^2}{m^2}\right)^{1/2}}\right) 
\end{equation}
with $P_{-n-1}=P_n$ and $P_n(1)=1$, making the approximation $\Delta/m \ll 1$ we see that 
\begin{equation}
_{2}F_{1} (-n, 1/2;1,-\Delta^2/m^2)=(1+\frac{\Delta^2}{m^2})^{n/2}\left(P(1)+ \mathcal{O}\left(\frac{\Delta^4}{m^4}\right)\right).
\end{equation}
Therefore we get,
\begin{equation}
I=-\frac{L}{2a}\sum_{n=1}^{\infty} \left[\beta m \left(1+\frac{\Delta^2}{m^2}\right)^{1/4}\right]^{2n} \frac{1}{(2n)!} \frac{1}{2n} (-1)^n B_n 
\end{equation}
with $m\left(1+\frac{\Delta^2}{m^2} \right) ^{1/4}=\omega_0$
\begin{equation}
I=-\frac{L}{2a} \sum_{n=1}^{\infty} (\omega_0 \beta)^{2n} \frac{(-1)^n B_n}{(2n)!2n}
\end{equation}
with $x=i \beta \omega_0$
\begin{equation}
I=-\frac{L}{2a} \sum_{n=1}^{\infty} \frac{x^{2n}}{(2n)!} \frac{B_n}{2n}.
\end{equation}
Taking the derivative with respect to $x$ and multiplying with $x$; we get,
\begin{equation}
x \frac{dI}{dx}= -\frac{L}{2a} \sum_{n=1}^{\infty} \frac{x^{2n}}{(2n)!} B_n
\end{equation}
from the identity in \cite{whittaker1996course} 
page 125,
\begin{equation}
\frac{x}{2} \cot(\frac{x}{2})=1-\sum_{n=1}^{\infty} \frac{x^{2n}}{(2n)!} B_n
\end{equation}
with $x=iz$; we have, 
\begin{equation}
\frac{dI}{dz}= -\frac{L}{2a} \left[\frac{1}{z}-\frac{1}{2}\coth(\frac{z}{2})\right].
\end{equation}
Integrating over $z$ with $z=\omega_0$
\begin{equation}
I=-\frac{L}{2a} \left[\log(z)-\log(\sinh(\frac{z}{2}))\right] +C_o
\end{equation}
where $C_0$ is the constant of integration.\par
Since $I(0)=0$
\begin{equation}\label{eq33}
I(\beta)=-\frac{L}{2a} \left[\log(\beta \omega_0)-\log(2\sinh(\frac{\beta \omega_0}{2}))\right].
\end{equation}

We now calculate the contribution of the second term in eq. (\ref{eq8}) to the series expansion.\par
We have defined $K$ such that (\ref{eq21}),
\begin{equation}
K=\frac{L}{\pi a} \int_{m}^{\sqrt{\Delta^2+m^2}} \frac{\omega \log(\beta \omega)d\omega}{((\omega^2-m^2)(\Delta^2+m^2-\omega^2))^{1/2}}.
\end{equation}
Setting $\omega^2=u$,
\begin{equation}\label{eq35}
\begin{split}
K=&\frac{L}{2 \pi a}\left\lbrace\int_{m^2}^{\Delta^2+m^2}\frac{\log(\beta)du}{\left[(u-m^2)(\Delta^2+m^2-u)\right]^{1/2}}\right.\\[8pt]
&\left. +\frac{1}{2} \int_{m^2}^{\Delta^2+m^2} \frac{\log(u)du}{\left[(u-m^2)(\Delta^2+m^2-u)\right]^{1/2}}\right\rbrace.
\end{split}
\end{equation}

We have therefore $K=K_1+K_2$ where $K_1\,(K_2)$ is the first (second) term in the above expression.

For the first term, $K_1$ we refer the reader Appendix C. $K_1=\frac{L}{2a}\log(\beta)$. For the second term on the right $(K_2)$, setting $u=m^2+\Delta^2-y$ we have
\begin{equation}
\begin{split}
K_2&=\frac{L}{4\pi a}\left\lbrace \int_{0}^{\Delta^2} \frac{dy \log(m^2+\Delta^2)}{[y(\Delta^2-y)]^{1/2}}\right. \\[8pt]
&\left.\quad+\int_{0}^{\Delta^2} \frac{dy\log(1-\frac{y}{m^2+\Delta^2})}{[y(\Delta^2-y)]^{1/2}}\right\rbrace.
\end{split}
\end{equation} 
The first term above again gives (Appendix C),
\begin{equation}\label{eq37}
=\frac{L}{2a} \log((m^2+\Delta^2)^{1/2}).
\end{equation}
Setting $y=\Delta^2z^2$, the second term gives,
\begin{equation}
=\frac{L}{2\pi a} \int_{0}^{1} \frac{dz\log(1-\frac{\Delta^2z^2}{m^2+\Delta^2})}{[1-z^2]^{1/2}}.
\end{equation}
From \cite{gradshteyn2014table} and Appendix $[D]$
\begin{equation}
K_2= \frac{L}{2a} \log(\frac{m+(\Delta^2+m^2)^{1/2}}{2(\Delta^2+m^2)^{1/2}}).
\end{equation} 
Finally, $K=K_1+K_2$ equals
\begin{equation}\label{eq30}
K=\frac{L}{2a} \log(\frac{\beta ((m^2+\Delta^2)^{1/2}+m)}{2}).
\end{equation}
Collecting results from (\ref{eq33}) and (\ref{eq30}) 
\begin{equation}
\begin{split}
\log(Z) &=\frac{L}{2a} \left\lbrace\log(\frac{\omega_0}{kT})-\log(2\sinh(\frac{\omega_0}{2kT}))\right. \\[8pt]
&  \left.\quad-\log(\frac{(m^2+\Delta^2)^{1/2}+m}{2kT})\right\rbrace
\end{split}
\end{equation}
with $\omega_0=m(1+\frac{\Delta^2}{m^2})^{1/4}$. Up to order $\frac{\Delta^4}{m^4}$
\begin{equation}
\log(Z)=-\frac{L}{2a} \log(2\sinh(\frac{\omega_0}{2kT})).
\end{equation} 

\subsection{Ground State Energy}
\hspace{\parindent}Now we calculate the ground state energy for our system of BSHOs. We have \begin{equation}
E_0=\sum_{k=0}\frac{\omega_k}{2}
\end{equation}
setting $\hbar=1$.

Using the expression for the enumeration of the momentum states
\begin{equation}
=\frac{L}{2\pi a}\int_{m}^{(\Delta^2+m^2)^{1/2}} \frac{\omega^2 d\omega}{((\omega^2-m^2)(\Delta^2+m^2-\omega^2))^{1/2}}.
\end{equation}
Setting $\omega^2=m^2+y$ and $2\omega d\omega=dy$
\begin{equation}\label{eq48}
E_0=\int_{0}^{\Delta^2}\frac{L}{4 \pi a} \frac{(m^2+y)^{1/2}dy}{(y(\Delta^2-y))^{1/2}}
\end{equation}
again using the result from \cite{gradshteyn2014table} and Appendix [A], with
\begin{equation}\label{eq46}
E_0=\frac{mL}{4 a}\: _{2}F_{1}\,[-1/2, 1/2; 1, -\Delta^2/m^2].
\end{equation}
Now we use the result in Appendix [B], that is \cite{abramowitz1965handbook} page 561,
\begin{equation}\label{eq47}
_{2}F_{1}\,[-1/2, 1/2; 1, -\Delta^2/m^2]= \left(1+\frac{\Delta^2}{m^2}\right)^{1/4}\,P_{-3/2}\left[\frac{(1+\frac{\Delta^2}{2m^2})}{(1+\frac{\Delta^2}{m^2})^{1/2}}\right]
\end{equation}
with $P_{1/2}=P_{-3/2}$ and $\Delta^2/m^2\ll1$, and $P_n(1)=1$ we have
\begin{equation}
E_0=\frac{mL}{4a} \left(1+\frac{\Delta^2}{m^2}\right)^{1/4}.
\end{equation}

\subsection{Thermodynamics of BSHOs}

We have for $Z$ from above
\begin{equation}
\log(Z)=-\frac{L}{2a}\log(2\sinh(\frac{\omega_0}{2kT}))
\end{equation}
due to the well known relationship $E=-\left(\frac{\partial\log(Z)}{\partial\beta}\right)_L$
\begin{equation}\label{eq40}
E=\frac{\omega_0L}{4a}\coth(\frac{\omega_0}{2kT}).
\end{equation}
Now for $\frac{\omega_0}{2kT}\ll1$
\begin{equation}
E=\frac{L}{2a}kT
\end{equation}
for $\frac{\omega_0}{2kT}\gg1$ 
\begin{equation}
E=\frac{\omega_0 L}{4a}
\end{equation}  
we get the same result for the ground state energy as in (2.4).

Now we will calculate the entropy of our system of BSHOs. Again we have to order of $\frac{\Delta^4}{m^4}\ll1$
\begin{equation}
A=\frac{L}{2a}kT \log(2\sinh(\frac{\omega_0}{kT}))
\end{equation}
with
\begin{equation}
S=-\left(\frac{\partial A}{\partial T}\right)_L=\frac{L}{2a}k\left(\frac{\omega_0}{2kT} \coth(\frac{\omega_0}{2kT})-\log(2\sinh(\frac{\omega_0}{2kT}))\right).
\end{equation}

Now we consider asymptotic behavior of entropy as $T\rightarrow0$, and $T\rightarrow \infty$. As $T\rightarrow0$, $S$ also goes to zero. Thus our expression for the entropy satisfies third law of thermodynamics.

As $T\rightarrow \infty$,
\begin{equation}
S=\frac{L}{2a}k \left(\log(\frac{kT}{\omega_0})+1\right).
\end{equation}

We have obtained an expression for $S, A, E$ in terms of $T$ and $L$. We take the variable $L$ for the volume of the system. Indeed in the case of 2d (3d) systems, (ref.MG unpublished) we have $L^2$ $(L^3)$ respectively which is consistent with our choice. Also naturally $P=-\left(\frac{\partial A}{\partial L}\right)_T$ which follows from the expression given for the free energy. Also we have $E+PL-TS=\mu N$. Since the number of phonons are not conserved $\mu=0$, $N$ is irrelevant and $E+PL-TS=0$. This is satisfied when the expressions for $E,S$ and $P$ are substituted into this expression. Thus we have everything is as it should be. 
 
 Following the discussion above we can define $L/2a$ to be the volume. Now we want to calculate explicitly the pressure of this model in this case. We have to order $\Delta^4/m^4$ using the Maxwell's relationship, $P=-\left(\frac{\partial A}{\partial L}\right)_T$ as noted above,
\begin{equation}
P=-\frac{kT}{2a} \log(2\sinh(\frac{\omega_0}{2kT})).
\end{equation}

As $T\rightarrow 0$, P goes to $-\frac{\omega_0}{4a}$. As $T\rightarrow \infty$, P goes to $\frac{kT}{2a}\log(\frac{kT}{\omega_0})$.Or more generally we have
\begin{equation}
P=-\frac{\omega_0}{4a}-\frac{kT}{2a}\log(1-e^{-\frac{\omega_0}{kT}}).
\end{equation}

We see that pressure can attain negative values. The temperature at $P=0$, is such that $2\sinh(\frac{\omega_0}{2kT_0})=1$, or  $\cosh(\frac{\omega_0}{2kT_0})=\frac{\sqrt{5}}{2}$.

Now we consider variations in pressure as a function of $T_c$, $\delta T=T-T_c$, $\delta P=P-P_c$. So we have 
\begin{equation}
\delta P=\frac{k}{4a} \left(\frac{\omega_0}{kT_c}\right) \sqrt{5} \:\delta T.
\end{equation}

We see that there exists a remote possibility of considering this behavior as a phase transition  such that $\delta P$ is an order parameter with the critical exponent $\alpha=1$, although the other thermodynamical functions are well defined and therefore lack examples of singular behavior to merit being called phase transitions.

Now we calculate specific heat at constant volume and constant pressure. From above (\ref{eq40}),
\begin{equation}
C_L=\left(\frac{\partial E}{\partial T}\right)_L=\frac{kL}{2a}\left(\frac{\omega_0}{2kT}\right)^2 \csch[2](\frac{\omega_0}{2kT}).
\end{equation}

As $T\rightarrow 0$, $C_L\rightarrow \frac{L}{2a} \left(\frac{\omega_0}{2kT}\right)^2e^{\frac{-\omega_0}{kT}}$ and as $T\rightarrow \infty$, $C_L=\frac{L}{2a}k$. We observe the ideal gas behavior in this formula. As $T\rightarrow 0$ on the other hand we observe a Schottky type anomaly in the specific heat.

Now we use the result from \cite{huang2008statistical} 
\begin{equation}
C_P-C_L=\frac{TL \alpha^2}{K_T}
\end{equation} 
where $\alpha=\frac{1}{L} \left(\frac{\partial L}{\partial T}\right)_P$ is the coefficient of linear expansion and $K_T=-\frac{1}{L}\left(\frac{\partial L}{\partial P}\right)_T$ is called the compressibility. Since $P$ is independent of $L$ we have, $\left(\frac{\partial P}{\partial L}\right)_T=0$ and $\frac{1}{K_T}=0$.

Since $P=P(T)$ only from above, so constant pressure means constant temperature i.e, $\left(\frac{\partial L}{\partial T}\right)_P$ is not defined.

Therefore, we make the conjecture $C_P=C_L$. We also see that thermodynamical relationship $E-TS+PL=\mu N$ is obeyed with $\mu N=0$ since number of phonons are not conserved. Therefore, our calculation is self-consistent.

\section{Calculation of the Partition Function of Fermionic Oscillators}
\subsection{The Partition Function}
\hspace{\parindent}In this section we consider the case of a system of fermionic oscillators (FSHO) with the same dispersion relationship as the previously given one.

In analogy with the previous calculation, we make the passage from discrete case to the continuous case. Our dispersion relationship reads the same as the previous case
\begin{equation}
\omega_k^2=m^2 +\Delta^2\sin[2](\frac{k\pi a}{L})
\end{equation}
for a single FSHO, we have from \cite{nakahara2016geometry} the expression for the partition function,
\begin{equation}
Z_k=2\cosh(\frac{\beta \omega_k}{2})
\end{equation}
and for a system of FSHOs we have
\begin{equation}
Z=\prod_{k} Z_k.
\end{equation}

Now going explicitly to the continuum limit and enumerating the momentum states,
\begin{equation}
\Delta k=\frac{L}{\pi a} \frac{\omega \Delta \omega}{[(\omega^2-m^2)(\Delta^2+m^2-\omega^2)]^{1/2}}.
\end{equation}

In the continuum limit 
\begin{equation}
\log(Z)=\frac{L}{\pi a}\int_{m}^{\sqrt{\Delta^2+m^2}} \frac{\omega d\omega}{[(\omega^2-m^2)(\Delta^2+m^2-\omega^2)]^{1/2}}\log(2\cosh(\frac{\beta \omega}{2}))
\end{equation}
expanding $\log$ term, in power of $e^{-\beta \omega}$ we have
\begin{equation}
\begin{split}
\log(2\cosh(\frac{\beta \omega}{2})) &=\frac{\beta \omega}{2}+\sum_{n=1}^{\infty} (-1)^{n+1}\, \frac{e^{-n \beta \omega}}{n}\\
&=\frac{\beta \omega}{2}+f_1(e^{-\beta\omega})
\end{split}
\end{equation}
where $f_1$ is the Fermi-Dirac integral function, of order 1, in direct analogy to the Bose-Einstein integral function.

We have also from \cite{pathria2011statistical}
\begin{equation}
f_1(e^{-\beta\omega})=g_1(e^{-\beta\omega})-g_1(e^{-2\beta\omega})
\end{equation}
where $g_1(e^{-\beta\omega})$ is the Bose-Einstein integral function of order 1.

For $g_1(e^{-\beta\omega})$ and $g_1(e^{-2\beta\omega})$ we have from \cite{pathria2011statistical},
\begin{equation}
g_1(e^{-\beta\omega})=\left[-\log(\beta\omega)+\sum_{i=1}^{\infty} (-1)^i \frac{B_i(\beta\omega)^{2i}}{(2i)!(2i)}+\frac{\beta\omega}{2}\right]
\end{equation}
\begin{equation}
g_1(e^{-2\beta\omega})=\left[-\log(2\beta\omega)+\sum_{i=1}^{\infty} (-1)^i \frac{B_i(2\beta\omega)^{2i}}{(2i)!(2i)}+\frac{2\beta\omega}{2}\right]
\end{equation} 
\begin{equation}\label{eq72}
f_1(e^{-\beta\omega})+\frac{\beta\omega}{2}=\left[\log(2)+\sum_{i=1}^{\infty}(-1)^i \frac{B_i(1-2^{2i})(\beta\omega)^{2i}}{(2i)!(2i)} \right]
\end{equation}
take the integral over $\omega$ with the density of states given above. Considering the first term in the parenthesis in (\ref{eq72}) and taking the integral over $\omega$ using our dispersion relationship we have as in the first part of (\ref{eq35}), defining $S_1$,
\begin{equation}
S_1=\frac{L}{\pi a}\log(2)\int_{m}^{\sqrt{\Delta^2+m^2}}\frac{\omega d\omega}{[(\omega^2-m^2)(\Delta^2+m^2-\omega^2)]^{1/2}}
\end{equation} 
from Appendix C, 
\begin{equation}
S_1=\frac{L}{2a}\log(2).
\end{equation}
For the second term on the other hand, defining $S_2$, 
\begin{equation}
S_2= \frac{L}{\pi a} \int_{m}^{\sqrt{\Delta^2+m^2}} \sum_{i=1}^{\infty} (-1)^i\frac{B_i (1-2^{2i})}{(2i)! 2i} (\beta \omega)^{2i} \frac{\omega d\omega}{(m^2-\omega^2)(\Delta^2+m^2-\omega^2))^{1/2}}.
\end{equation}
As a result we have 
\begin{equation}
S_2=I(\beta\omega_0)-I(2\beta\omega_0)
\end{equation}
where $I$ is defined in (\ref{eq20}) and in (\ref{eq33}). Thus, $S=S_1+S_2$ and
\begin{equation}
\begin{split}
\log(Z)&=\frac{L}{2a}\left\lbrace \left[\log(\beta\omega_0)-\log(2\sinh(\frac{\omega_0\beta}{2}))\right]\right.\\[8pt]
&\left.-\left[\log(2\beta\omega_0)-\log(2\sinh(\omega_0\beta))\right]+\log(2)\right\rbrace.
\end{split}
\end{equation}
Finally we have
\begin{equation}
\log(Z)=\frac{L}{2a}\log(2\cosh(\frac{\beta\omega_0}{2})).
\end{equation}

\subsection{Thermodynamical Quantities for FSHOs}
\hspace{\parindent}We consider the quantities $S,P,E,L,A,C_L,C_P$ (entropy, pressure, energy, volume, Helmholtz free energy, specific heat at constant volume, specific heat at constant pressure respectively).

We have from above 
\begin{equation}
\log(Z)=\frac{L}{2a} \log(2\cosh(\frac{\omega_0}{2kT}))\\
\end{equation}
\begin{equation}
A=-kT\log(Z)
\end{equation}
\begin{equation}
E=-\left(\frac{\partial\log(Z)}{\partial\beta}\right)_L=-\frac{L}{4a} \omega_0 \tanh(\frac{\omega_0}{2kT})
\end{equation}
with
\begin{equation}
C_L=\left( \frac{\partial E}{\partial T}\right)_L=\frac{L}{2a} \left(\frac{\omega_0}{2kT}\right)^2 \sech[2](\frac{\omega_0}{2kT})k.
\end{equation}

In direct analogy to the results obtained in (2.5) and the thermodynamical identity from \cite{huang2008statistical} 
\begin{equation}
C_P-C_L=\frac{TL\alpha^2}{K_T}
\end{equation}
again since $\left(\frac{\partial P}{\partial L}\right)_T=0$, $1/K_T=0$, $\alpha$ is not defined, so we say that $C_P=C_L$.

We now give the expression for $S$, the entropy
\begin{equation}
S=-\left(\frac{\partial A}{\partial T}\right)_L=\frac{L}{2a}k\log(2\cosh(\frac{\omega_0}{2kT}))-\frac{\omega_0}{2kT}\tanh(\frac{\omega_0}{2kT}).
\end{equation}
Next we consider the asymptotic behavior as $T\rightarrow 0$ and $T\rightarrow \infty$ for the thermodynamical quantities above. First we consider $E$ when temperature goes to zero,
\begin{equation}
E=-\frac{L}{4a}\,\sgn(T)
\end{equation}
and as the temperature goes to infinity,
\begin{equation}
E=\frac{L}{2a} \left(\frac{\omega_0}{2kT}\right)\frac{\omega_0}{2}.
\end{equation}
Now consider the behavior of specific heat at constant volume
\begin{equation}
\text{as}\quad T\rightarrow 0_+\qquad C_L=\frac{L}{2a}\left(\frac{\omega_0}{2kT}\right)^2 e^{-\left(\frac{\omega_0}{2kT}\right)}
\end{equation}
\begin{equation}
\text{as}\quad T\rightarrow 0_-\qquad C_L=\frac{L}{2a}\left(\frac{\omega_0}{2kT}\right)^2 e^{\left(\frac{\omega_0}{2kT}\right)}.
\end{equation}
Now we have for $S$, the asymptotic behavior as $T\rightarrow \pm\infty$ as given below, 
\begin{equation}
S=k\frac{L}{2a}\left[\log(2)-\left(\frac{\omega_0}{2kT}\right)^2\right]
\end{equation}
As $T\rightarrow 0$ on the other hand,
\begin{equation}
S=0.
\end{equation}

We see that our equation for entropy satisfies the third law of thermodynamics. Also, for $T$ goes to infinity or minus infinity, the entropy goes as $\log(2)$ which is the entropy of a system of paramagnetic material as $T$ goes to infinity or minus infinity.\par
From above we see immediately that $T$ can attain negative values as a result of the dependence of $A,S,E$ on temperature as given by for example \cite{pathria2011statistical}. Indeed the thermodynamical quantities for a system of FSHOs, $A,E, S$ are exactly the same as the ones given in the case of a paramagnetic system of spin one half dipoles in the presence of a magnetic field, with $\frac{L}{2a}$ exchanged with $N$. Since the paramagnetic system can attain negative values of temperature, our system can also do so. The reason for the paramagnetic system to obtain negative values is that the energy is bounded from above. Bosonic systems similar to the one which is considered in this paper are not bounded from above and therefore do not attain negative temperatures.

Finally we will calculate the pressure in terms of $L$ and $T$. We have for our system of FSHOs a well known expression for the Helmholtz free energy and derivative with respect to $L$. First we let $\frac{L}{2a}$ to stand for volume. Then, with 
\begin{equation}
A=-kT \frac{L}{2a} \log(2\cosh(\frac{\omega_0}{2kT}))
\end{equation}
\begin{equation}
P=-\left(\frac{\partial A}{\partial L}\right)_T=\frac{kT}{2a}\log(2\cosh(\frac{\omega_0}{2kT})).
\end{equation}

Now we consider the asymptotic behavior of the pressure $P$, as $T$ goes to zero and as $T$ goes to plus or minus infinity. As $T$ goes to zero,
\begin{equation}
P=\frac{\omega_0}{4a} \sgn(T)\quad \text{where}\quad \sgn(T)=\theta(T)-\theta(-T)
\end{equation} 
we observe discontinuous behavior of pressure as $T$ goes to zero from above and from below. Now as $T$ goes to infinity,
\begin{equation}
P=\frac{kT}{2a}\left(\log(2)+\frac{1}{2}\left(\frac{\omega_0}{2kT}\right)^2\right).
\end{equation}

\section{Conclusion}
\hspace{\parindent}We have shown that for our one dimensional systems of BSHOs and FSHOs energy and specific heat as a function of the temperature $T$, lattice size $L$, lattice spacing $a$ are given by 
\begin{align}
\text{FSHOs}\quad E&=-\frac{L}{4a}\omega_0 \tanh(\frac{\omega_0}{kT})&& C_L=\frac{L}{2a}\left(\frac{\omega_0}{2kT}\right)^2\sech[2](\frac{\omega_0}{2kT})\\[8pt]
\text{BSHOs}\quad E&=-\frac{L}{4a}\omega_0 \coth(\frac{\omega_0}{kT})&& C_L=\frac{L}{2a}\left(\frac{\omega_0}{2kT}\right)^2\csch[2](\frac{\omega_0}{2kT}).
\end{align}
We note that for the FSHOs we see that there is a discontinuity in $E$ as a function of temperature i.e. 
\begin{align}
T\rightarrow 0_+ && E&=-\frac{L}{4a}\omega_0\\
T\rightarrow 0_- && E&=\frac{L}{4a}\omega_0.
\end{align}
We discussed the behavior of pressure for FSHOs which is given by 
\begin{equation}
P=\frac{kT}{2a}\log(2\cosh(\frac{\omega_0}{2kT}))
\end{equation}
as $T$ goes to zero 
\begin{equation}
P=\frac{\omega_0}{4a} \sgn(T).
\end{equation}

Thus we see that in the case where $L$ stands for the volume of the system there is a discontinuity in $E$ and $P$ at $T=0$. This does not pose any problem however since the real thermodynamical parameter is $1/T$ and not $T$. Therefore $T=0_-$ and $T=0_+$ are infinitely apart from each other. Hence there is no real discontinuity. No such behavior is observed for BSHOs. For the bosonic simple harmonic case with
\begin{equation}
P=-\frac{kT}{2a}\log(2\sinh(\frac{\omega_0}{2kT}))
\end{equation}
we see that for $P=0$, we have for $T_C$ 
\begin{equation}
kT_C=\frac{\omega_o}{\log(\frac{5}{4})}
\end{equation}
for $T<T_C$, $P<0$ is only (looks) like a phase transition.

Finally as far as entropy is concerned
\begin{equation}
S=\frac{L}{2a}k\left(\left(\frac{\omega_0}{2kT}\coth(\frac{\omega_0}{2kT})\right)-\log(2\sinh(\frac{\omega_0}{2kT}))\right)
\end{equation}
for BSHOs and
\begin{equation}
S=\frac{L}{2a}k\left(\log(2\cosh(\frac{\omega_0}{2kT}))-\frac{\omega_0}{2kT}\tanh(\frac{\omega_0}{2kT})\right)
\end{equation}
for FSHOs. 

We see immediately that $S$ is a continuous function of $T$ and $L$ for both FSHOs and BSHOs. For $T$ goes to $0_+$ $S=\frac{L}{2a}k\log(2)$ (identically the same as the paramagnetic system of spin 1/2 dipoles in the presence of a magnetic field).

As noted above while attempting to calculate $C_P$ there is a problem with the well known equation \cite{huang2008statistical} $C_P-C_L=\frac{\alpha^2}{K_TT}$ where $\alpha$ is the coefficient of thermal expansion
\begin{equation}
\alpha=\frac{1}{L}\left(\frac{\partial L}{\partial T}\right)_P
\end{equation} 
and the compressibility
\begin{equation}
K_T=\frac{1}{L}\left(\frac{\partial L}{\partial P}\right)_T
\end{equation}
$\frac{1}{K_T}$ equals to zero, since $\left(\frac{\partial P}{\partial L}\right)_T$ vanishes. On the other hand with $\alpha=\frac{1}{L}\left(\frac{\partial L}{\partial T}\right)_P$ since $P$ is a function of temperature only and because when we take derivative with respect to temperature we have to keep $P$ fixed $\alpha$ is not defined.

Another thing which comes up is that for a single oscillator 
\begin{equation}
Z=\log(2\sinh(\frac{\omega}{2kT}))\left(\log(2\cosh(\frac{\omega}{2kT}))\right)
\end{equation}
for BSHO (FSHO) where $\omega$ is the eigenfrequency of the Hamiltonian. For our case however we have a collection of BSHOs (FSHOs) and the total result after summing over eigenmodes is that 
\begin{equation}
Z=-\frac{L}{2a}\log(2\sinh(\frac{\omega_0}{2kT}))\left(\log(2\cosh(\frac{\omega_0}{2kT}))\right).
\end{equation}

For bosonic and fermionic cases seperately we also have $\omega_0=m\left(1+\frac{\Delta^2}{m^2}\right)^{1/4}$ and if $L$ is taken to be the number of oscillators we have that $\omega_0$ is something like the average of frequencies summed over with modes.

Apart from the discontinuity in $E$ and $P$ as $T$ goes to zero for FSHOs and from the behavior of BSHOs such that $P_C<0$ for $T<T_C$ all the physical quantities are well behaved functions of temperature and volume. Thus we can tentatively state that we will not observe a traditional phase transition. We have only made one approximation (i.e. $\Delta/m\ll 1$), otherwise our results are valid for all $T$ and $L$.

Universality demands that no phase transition is possible for $\Delta/m \approxeq 1$ or $\Delta/m \gg 1$, if there is no phase transition for $\Delta/m \ll 1$ since the form of the Hamiltonian remains the same.

One amusing thing which comes up above is that with the standard relationship for the thermodynamical quantities we have the formula
\begin{equation}
E-TS+PV=\mu N
\end{equation}

However for our system we are making calculations with phonons, the numbers of which are not conserved so $N$ is not a parameter and $\mu$ is defined to be zero.

In the two cases which we have studied above this relation is satisfied for the case $L/2a$ stands for the volume and $P$ is thus defined.

\section{Acknowledgements}The authors would like to thank the physics department of Boğaziçi University for creating a friendly atmosphere for research. The authors would also like to thank Ş. Şahin for her diligent effort in typing this paper. This work was supported in part by a fellowship from Turkish Academy of Sciences.

\bibliographystyle{abbrv}
\bibliography{refs} 

\appendix
\numberwithin{equation}{section}

\section{Certain Relationships Involving Hypergeometric Functions}
\hspace{\parindent}We want to solve our integral given above in (\ref{eq23}) for case 1 
\begin{equation}
I_n=\frac{L}{2\pi a }\frac{1}{2} \int_{0}^{\Delta^2} \frac{(m^2+x)^n dx}{(x (\Delta^2-x))^{1/2} }.
\end{equation}
Now from \cite{gradshteyn2014table} we have 
\begin{equation}
\int_{0}^{u} dx\, x^{\nu-1} (x+a)^\lambda (u-x)^{\mu-1}=\alpha^\lambda u^{\mu+\nu-1} B(\mu,\nu)\,_{2}F_{1}\,[-\lambda, \nu, \mu+\nu,-u/\alpha]
\end{equation}
where $_{2}F_{1}$ is a hypergeometric function and $B(\mu,\nu)$ is the beta function. Substituting in $\mu=\nu=1/2$, $\lambda=n$, $\alpha=m^2$ and $u=\Delta^2$ and with $B(1/2,1/2)=\pi$, 
\begin{equation}
I_n=\frac{L}{2a} m^{2n}\, _{2}F_{1}[-n,1/2;1,-\Delta^2/m^2]
\end{equation}
Now for case 2, equation (\ref{eq48}) with the choice of parameters $\mu=\nu=1/2$, $\lambda=1/2$, $\alpha=m^2$ and $u=\Delta^2$ and with $B(1/2,1/2)=\pi$ gives us equation (\ref{eq46}).
\section{Certain Relationships Involving Legendre Polynomial}
\hspace{\parindent}We have from \cite{abramowitz1965handbook} page 561, for the first case, i.e. (\ref{eq26})
\begin{equation}
F(a,b;2b,z)=2^{2b-1}\,\Gamma\left(\frac{1}{2}+b\right) z^{1/2-b}(1-z)^{\frac{1}{2}\left(b-a-\frac{1}{2}\right)}
P_{a-b-1/2}^{1/2-b}\left(\frac{1-\frac{z}{2}}{(1-z)^{1/2}}\right)
\end{equation}
where $P_\mu^\nu$ is the Legendre function. 

Now we set $a=-n$, $b=1/2$ and $z=-\Delta^2/m^2$ and we get
\begin{equation}
_{2}F_{1}[-n,1/2;1,-\Delta^2/m^2]=\left(1+\frac{\Delta^2}{m^2}\right)^{m/2} P_{-n-1}\left[\frac{1+\frac{\Delta^2}{2m^2}}{\left(1+\frac{\Delta^2}{m^2}\right)^{1/2}}\right]
\end{equation}
which is a Legendre function, and now using the well known properties of this function such that 
\begin{equation}
P_{-n-1}(z)=P_{n}(z),\qquad P_{n}(1)=1\quad \text{for all n}.
\end{equation}
Now
\begin{equation}
P_{n}\left[\frac{1+\frac{\Delta^2}{2m^2}}{\left(1+\frac{\Delta^2}{m^2}\right)^{1/2}}\right]=P_{n}\left(1+\mathcal{O}\left(\frac{\Delta^4}{m^4}\right)\right)=1
\end{equation}
for $\Delta^2/m^2\ll 1$. This is the only approximation we make and the net result is
\begin{equation}
_{2}F_{1}[-n,1/2;1,-\Delta^2/m^2]=\left(1+\frac{\Delta^2}{m^2}\right)^{n/2}.
\end{equation} 
Now for the second case, i.e. (\ref{eq47})
\begin{equation}
_{2}F_{1}\,[-1/2, 1/2; 1, -\Delta^2/m^2]= \left(1+\frac{\Delta^2}{m^2}\right)^{1/4}\,P_{-3/2}\left[\frac{(1+\frac{\Delta^2}{2m^2})}{(1+\frac{\Delta^2}{m^2})^{1/2}}\right]
\end{equation}
where we choose $a=-1/2$, $b=1/2$ and $z=-\Delta^2/m^2$.

\section{An Integral Arising from an Algebraic Expression as the Integrand}
\hspace{\parindent}Our integral is for the first case (\ref{eq35})
\begin{equation}
K=\frac{L}{2\pi a}\int_{m^2}^{\Delta^2+m^2} \frac{dx \log(\beta)}{[(x-m^2)(m^2+\Delta^2-x)]^{-1/2}}.
\end{equation}
From \cite{gradshteyn2014table} page 285, 
\begin{equation}
\int_{a}^{b} (x-a)^{\mu-1} (b-x)^{\mu-1}dx=(b-a)^{\mu+\nu-1} B(\mu,\nu)
\end{equation}
where $B(\mu,\nu)$ is the beta function. We have $\mu=\nu=1/2$, as a result
\begin{equation}
K=\frac{L}{2a}\log(\beta).
\end{equation}
For the second case, (\ref{eq37}), $\mu=\nu=1/2$, $a=0$, $b=\Delta^2$.
\section{An Integral Where a Certain Logarithmic Integrand is Involved}
\hspace{\parindent}We have the integral
\begin{equation}
K_0'=\frac{L}{2\pi a}\int_{0}^1 dx \frac{\log(1-\frac{\Delta^2 x^2}{m^2 +\Delta^2})}{(1-x^2)^{1/2}}.
\end{equation}
From \cite{gradshteyn2014table} 4.295.38
\begin{equation}
\int_0^1dx\frac{\log(1+ax^2)}{(1-x^2)^{1/2}}=\pi \log(\frac{1+(1+a)^{1/2}}{2})\quad a\geq-1 .
\end{equation}
Now we set $a=-\frac{\Delta^2}{\Delta^2+m^2}$. Finally we get 
\begin{equation}
K_0'=\frac{L}{2a} \log(\frac{m+(m^2+\Delta^2)^{1/2}}{2(m^2+\Delta^2)^{1/2}}).
\end{equation}

\end{document}